\documentclass[12pt]{article}
\usepackage[utf8]{inputenc}
\usepackage[margin=1in]{geometry}
\usepackage{amsmath,amssymb}
\usepackage{graphicx}
\usepackage{booktabs}
\usepackage{titlesec}
\usepackage{microtype}
\usepackage{xcolor}
\usepackage{cite}
\usepackage{hyperref}

\hypersetup{colorlinks=true,linkcolor=blue,citecolor=red,urlcolor=blue}

\newcommand{\figref}[2][]{\textcolor{blue}{Fig.~\ref{#2}#1}}

\renewcommand{\thesection}{\Roman{section}}
\renewcommand{\thesubsection}{\Alph{subsection}}

\titleformat{\section}[block]
  {\centering\normalfont\bfseries}{\thesection.}{0.6em}{\MakeUppercase}
\titleformat{\subsection}[block]
  {\normalfont\itshape}{\thesubsection.}{0.6em}{}
\titleformat{\subsubsection}[block]
  {\normalfont\itshape}{}{0em}{}
\titlespacing{\section}{0pt}{2ex plus 1ex minus .2ex}{1.6ex}
\titlespacing{\subsection}{0pt}{1.8ex plus 1ex minus .2ex}{1.2ex}
\titlespacing{\subsubsection}{0pt}{1.4ex plus 1ex minus .2ex}{0.8ex}

\title{\textbf{{\large Tuning spreading transport in GaN HEMTs with an AlN space layer under non-equilibrium heating}}}

\author{\normalsize Daosheng Tang,$^{1,2,3,*}$\quad Rui Hu,$^{1}$\quad Chen Ning,$^{5}$\quad Bingyang Cao,$^{4,*}$\\[0.7em]
\begin{minipage}{\dimexpr\textwidth-2\tabcolsep\relax}\centering\footnotesize\itshape
$^{1}$School of Rail Transportation, Soochow University, Suzhou 215131, China\\
$^{2}$Jiangsu Key Laboratory of Power Batteries for Manned Aviation, Suzhou 215131, China\\
\scalebox{0.95}{$^{3}$Intelligent Urban Rail Engineering Research Center of Jiangsu Province, Suzhou 215131, China}\\
\scalebox{0.9}{$^{4}$Key Laboratory for Thermal Science and Power Engineering of Ministry of Education,}\\
Department of Engineering Mechanics, Tsinghua University, Beijing 100084, China\\
$^{5}$Inspur Electronic Information Industry Co., Ltd, Jinan 250000, China\\
$^{*}$Corresponding authors: dstang@suda.edu.cn (D.T.); caoby@tsinghua.edu.cn (B.C.)
\end{minipage}}

\date{\normalsize (Dated: \today)}

\begin{document}
\maketitle

\begin{center}{\bfseries Abstract}\end{center}
\noindent
Lateral spreading transport dominates near-junction heat dissipation in GaN high-electron-mobility transistors, yet thermal control has largely overlooked this geometry. It neither addresses heat spreading within the core AlGaN/GaN region nor accounts for the spectral non-equilibrium of the heat source. Here we tune the spreading resistance directly by inserting a thin, highly conducting AlN space layer above the source to intercept heat and divert it laterally. We test this approach across GaN, AlGaN/GaN, and AlGaN/AlN/GaN structures under both equilibrium and longitudinal-optical (LO)-pumped non-equilibrium heating, pairing non-equilibrium molecular dynamics (NEMD) with full-band phonon Monte Carlo (MC). In equilibrium, the AlN layer lowers the spreading resistance by 38\% relative to the AlGaN-capped structure and by 17\% relative to bare GaN, with little change in the one-dimensional resistance. The equilibrium color map over AlN thickness and interfacial bonding shows a strong interface effect but a weak thickness trend that reverses, so the gain is set by interface admission rather than solely by thickness or bulk thermal conductivity. This ordering is independently confirmed by gray MC simulations. Under LO pumping, the spreading resistance rises by 58--83\% in NEMD and by 9--20\% in MC, and the equilibrium spatial optimum weakens or even reverses. This is because an optical-to-acoustic conversion gate now precedes every spatial gate and locks energy in slow optical modes. The three gates in series (optical-to-acoustic conversion, interface admission, and lateral diversion) close into a single non-equilibrium resistance network that predicts the full-band MC cases and frames the AlN space layer as a design variable within the non-equilibrium source state.

\vspace{1em}

\clearpage
\section{Introduction}

Gallium nitride (GaN) high-electron-mobility transistors (HEMTs) underpin high-power and high-frequency electronics through their wide bandgap \cite{1} and the high-density two-dimensional electron gas (2DEG) at the AlGaN/GaN heterointerface. As devices are driven harder and shrunk toward the nanometre scale, however, near-junction self-heating has become a primary limit on output power, gain, and reliability \cite{2}. The heat is concentrated at the drain-side gate edge over a region only tens of nanometres wide \cite{3}, and the flux is forced through this spatial constriction before reaching any remote sink, making the thermal spreading resistance a dominant part of the total. The same hotspot is also far from spectral equilibrium: Joule energy enters the lattice mainly through longitudinal-optical (LO) phonons \cite{4}, which must first convert to acoustic modes before the energy can spread, and localised Peltier--Thomson heat adds a further nanoscale contribution that can be tuned by doping \cite{5,6}. Near-junction heat is thus shaped simultaneously in space, by the constriction and its interfaces, and in the phonon spectrum, by optical-to-acoustic conversion, and the present work asks how the two can be controlled together.

These observations frame the engineering of near-junction heat removal. Early and still dominant strategies tune scalar coefficients---substrate thermal conductivity, diamond capping, interface conductance, and anisotropy---and treat the heat-flow pathway as fixed rather than as a design variable. The limits of this view are now clear: a top-side diamond layer is gated by the interface, and improving that interface is what lowers the resistance, confirming that admission into the high-thermal-conductivity body, not the thermal conductivity itself, is the bottleneck \cite{7,8}. The same interface gate appears across SiO$_2$--GaN, graphene/GaN, and MoS$_2$/GaN systems, where the transmission is tied to phonon-spectrum mismatch \cite{9,10}, while optimizing buffer thickness stays within the same layer-by-layer framework \cite{11}. A high thermal conductivity placed where heat cannot enter it therefore does not guarantee a low spreading resistance. Once the scalar view is recognized as insufficient, the spreading geometry itself becomes the design target, part of a broader move toward thermal design automation \cite{12}. Topology optimization redistributes high-thermal-conductivity material along the actual constriction rather than adding it uniformly \cite{13}, and a parallel lesson on placement shows that a high-thermal-conductivity layer embedded inside the heat-generating region intercepts and diverts heat laterally, whereas the same material at the far boundary only widens the back isotherm, so the two actions do not add. This motivates a thin, in-plane conducting layer at the heat-generating interface; earlier thermal studies treated AlN only through its through-thickness conductance \cite{14}, and back-barrier and AlN/GaN/AlN designs were characterized for their cross-plane resistance \cite{15,16}, leaving the lateral route open.

An AlN space layer is the natural choice for this lateral, high-thermal-conductivity route: it is highly conducting in-plane and supports long acoustic mean free paths, so it intercepts heat at the source and diverts it before the flow is constricted. The same layer is already a well-established part of AlGaN/GaN heterostructures for electrical reasons: it is grown coherently as a single crystal on the GaN lattice, raising the two-dimensional electron gas density without compromising mobility or carrier confinement \cite{17,18}, so its thermal and electrical functions are compatible and one layer can be tuned to serve both \cite{19}.

Whether this layer can deliver the intended lateral route, however, is governed by the transport physics of a confined nanoscale source. A single Fourier-based description is insufficient: classical thermal-resistance networks treat spreading as a geometry-dependent correction and remain accurate only while the source is large relative to the phonon mean free path (MFP) \cite{20}. Once the source width approaches the MFP, Fourier-based diffusion ceases to be self-contained, and phonon Monte Carlo (MC), semi-empirical, radial, and Boltzmann treatments all place the spreading resistance above the Fourier-based prediction, with the deviation growing with the Knudsen number \cite{21,22}, a ballistic--diffusive spreading since imaged directly in operating transistors \cite{23}. Full-band MC further resolves the resistance by phonon branch and shows it to depend on the source type \cite{24,25}, bias, and geometry \cite{26,27,28}, rather than on one effective thermal conductivity \cite{25}. The same finite, broadly distributed MFP also drives the size effect on thermal conductivity: a body smaller than the long MFPs, captured by effective-MFP models \cite{29}, cannot use its bulk value, and slip boundary conditions represent the long-MFP phonons that cross it without relaxing \cite{30}, a framework that extends to multiply-constrained nanostructures \cite{31}, to metallic and engineered analogues \cite{32,33}, and to the anisotropic, thickness-dependent thermal conductivity of real HEMT stacks \cite{34,35}. The spreading resistance and the size-dependent thermal conductivity therefore share one origin in the finite phonon MFP. The spatial redesign must also be coupled to the non-equilibrium source. Electrons shed energy selectively onto LO and interface-optical modes that decay slowly and, at high drive, overshoot the acoustic population to form the hot-phonon bottleneck, measured in lattice-matched InAlN/AlN/GaN \cite{36}, at GaN/AlN interfaces and superlattices \cite{37,38}, and near GaN hotspots \cite{39}; coupled electron--phonon MC likewise gives a population that no single temperature can represent \cite{40,41}. Whether a space layer can help thus depends on optical-to-acoustic conversion that lets energy reach the interface, and spatial and spectral effects, though coupled in one hotspot, are rarely treated within a single framework.

Yet inserting this layer also raises the questions that decide whether the route works. It introduces two interfaces, and phonons can enter only through their frequency-dependent transmission, so each interface acts as an admission gate rather than a passive boundary. Within a film this thin, the in-layer mean free path is set against the layer thickness itself, and the confined modes and their in-plane propagation differ from bulk AlN. Coherent growth, moreover, locks the layer into elastic strain from the lattice mismatch, which reshapes both the polarization and the phonon spectrum. The benefit is thus governed by coupled interface, in-layer, and strain effects rather than by the bulk thermal conductivity of AlN alone.

We therefore take the spreading resistance as the quantity to be controlled and examine it across three structures and two source states, pairing atomistic non-equilibrium molecular dynamics (NEMD), which resolves lattice strain, interface distortion, and bonding \cite{42}, with full-band spectral phonon MC as a mesoscopic reference \cite{25}, and apply both equilibrium and optically pumped, non-equilibrium sources to separate the roles of the real interface and the intrinsic transport law. Across three structures and two source types, this turns the spreading resistance from an observed geometric outcome into a tunable quantity. In equilibrium the AlN space layer lowers it selectively without trading off the one-dimensional resistance, with interface admission rather than thickness setting the gain. Counterintuitively, that very gain weakens and can reverse under LO pumping: the pump raises the spreading resistance by 58--83\% and the equilibrium optimum becomes the worst case, because optical-to-acoustic conversion now precedes every spatial gate. The three steps (optical-to-acoustic conversion, interface admission, and lateral diversion) then close into one non-equilibrium resistance network, providing a model for designing near-junction spreading rather than merely adding another high-thermal-conductivity material.

\section{Materials and computational methods}

\subsection{Device structures and interatomic potential}

\begin{figure}[htb]
\centering
\includegraphics[width=\textwidth]{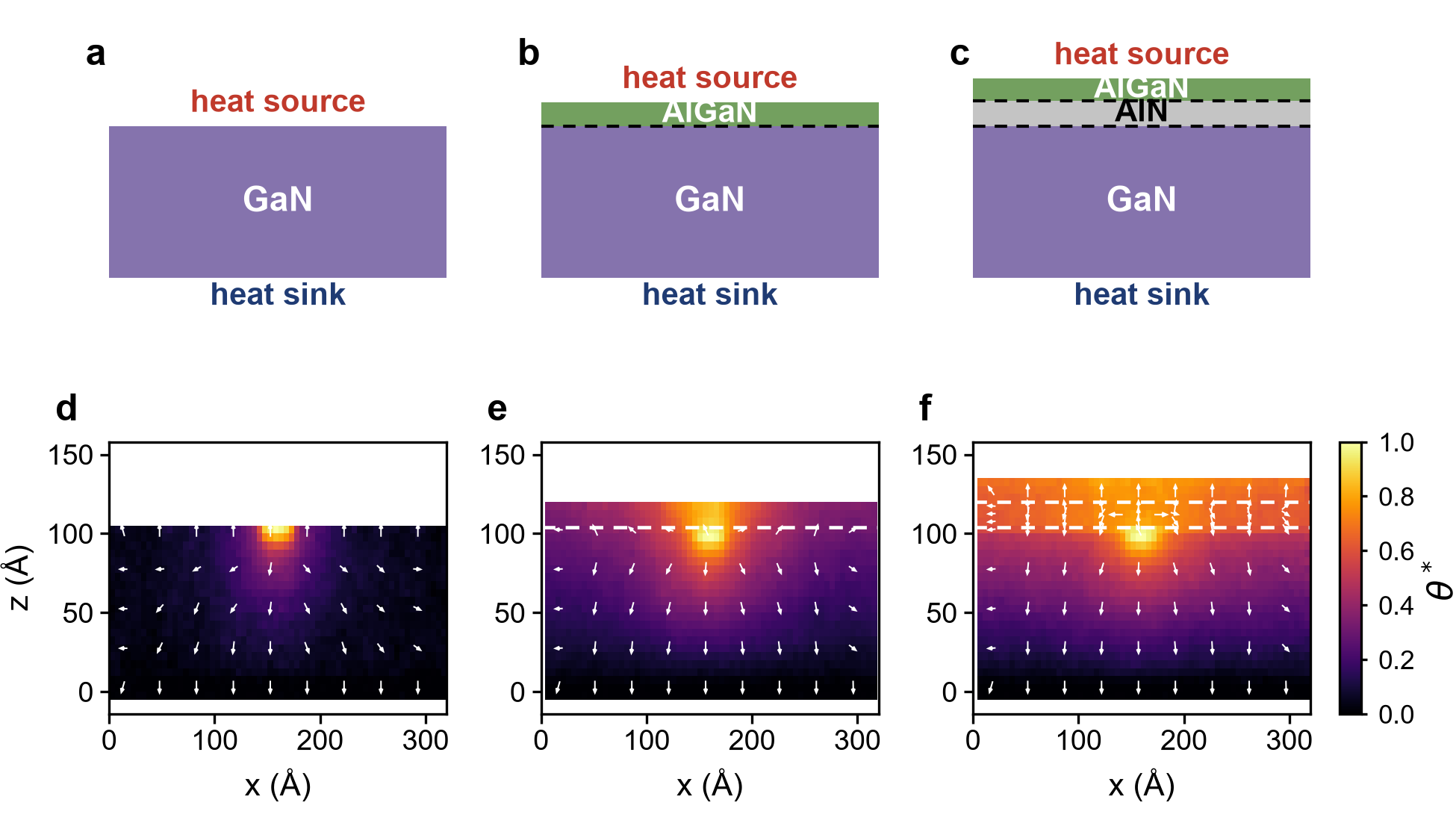}
\caption{Computational structures and near-junction thermal pathways. (a--c) Schematics of the three structures compared: (a) bare GaN, (b) AlGaN/GaN, and (c) AlGaN/AlN/GaN, with an identical GaN layer. Heat is generated in a finite slab inside the GaN active layer, removed at the bottom sink, and the top surface outside the source is adiabatic. (d--f) Corresponding normalized temperature fields $\theta^*=(T-T_{\mathrm{sink}})/(T_{\mathrm{max}}-T_{\mathrm{sink}})$, each mapped from 0 to 1. White arrows indicate the local heat-flow direction and are normalized to show direction only, not magnitude. The flow is strongly constricted in the bare GaN case; the AlGaN cap returns or traps upward-going phonons; the AlN space layer intercepts energy and diverts it laterally.}
\label{fig1}
\end{figure}

We examine three epitaxial structures that isolate the role of the AlN layer (\figref{fig1}): (i) a single-crystal GaN film as the reference (\figref[a]{fig1}); (ii) an AlGaN/GaN heterostructure (\figref[b]{fig1}); and (iii) an AlGaN/AlN/GaN composite in which a thin AlN space layer is inserted at the AlGaN/GaN interface, directly beneath the heat-generating region (\figref[c]{fig1}). Heat is released from a slab inside the GaN active layer and removed at the bottom face, while the top surface is held adiabatic.

The corresponding near-junction thermal response is shown in \figref[d--f]{fig1}. Each panel maps the normalized temperature $\theta^*=(T-T_{\mathrm{sink}})/(T_{\mathrm{max}}-T_{\mathrm{sink}})$ from 0 to 1, so the field patterns are compared directly rather than through their absolute temperatures. White arrows give the local heat-flow direction and are normalized to equal length, indicating direction only rather than magnitude. The flow is strongly constricted beneath the source in bare GaN, the AlGaN cap turns back or traps the upward-going component, and the AlN layer in the third structure intercepts energy at the interface and diverts it laterally, reflected by the horizontal arrows within the AlN layer.

Although this work addresses thermal transport, the thin AlN layer is a standard component of electrical AlGaN/GaN HEMTs. Its large conduction-band offset strengthens 2DEG confinement and suppresses alloy scattering, so in electrical devices ultrathin (sub-nanometre to about 1~nm) AlN layers increase the carrier concentration, mobility, and transconductance rather than degrade them \cite{43,44}. The higher barrier can increase ohmic contact resistance, but this is handled by contact recessing and is independent of the lateral spreading pathway examined here.

The atomistic cell is periodic in the two in-plane directions and free in the cross-plane direction, with lateral dimensions $31.9\times2.76$~nm. The relaxed layer thicknesses are $10.37$~nm of GaN, $1.56$~nm of AlN, and $1.29$~nm of Al$_{0.30}$Ga$_{0.70}$N, with the AlN and AlGaN stacked above the GaN. The mesoscopic MC devices use representative dimensions of $1.0~\mu$m GaN, $20$~nm AlGaN, and a $5$~nm AlN layer.

All atomistic interactions are described by a Tersoff bond-order potential \cite{45}, and the molecular-dynamics calculations are implemented in LAMMPS \cite{46},
\begin{equation}
E=\frac{1}{2}\sum_i\sum_{j\neq i} f_C(r_{ij})\left[f_R(r_{ij})+b_{ij}f_A(r_{ij})\right],
\end{equation}
where $f_R$ and $f_A$ are the repulsive and attractive pair terms, $f_C$ is the cutoff function, and the bond-order coefficient $b_{ij}$ depends on the bond lengths and angles to all neighboring atoms. A single parameter file is used consistently for GaN, AlN, and AlGaN; the Ga--N parameters are identical in the binary and ternary, so the GaN phase is physically equivalent in every structure. The potential reproduces the lattice constants, elastic moduli, and thermal conductivity of the binaries. After minimization, each structure is relaxed under isothermal--isobaric and then canonical conditions so the interfaces settle under their natural misfit strain, and region boundaries are defined from the relaxed atomic coordinates. The full NEMD protocol and the potential parameterization are detailed in \textcolor{blue}{Supplementary Note S1}.

\subsection{Thermal-resistance extraction}

For each structure the resistance is obtained from a two-source construction that separates lateral spreading from bulk one-dimensional conduction. A full-width source drives purely one-dimensional conduction and gives $R_{\mathrm{1D}}$, whereas a narrow source, one tenth of the film width, drives lateral spreading and gives the total resistance $R_{\mathrm{tot}}$. The spreading resistance is the difference,
\begin{equation}
R_{\mathrm{spread}}=R_{\mathrm{tot}}-R_{\mathrm{1D}}.
\end{equation}
The two simulations share the same materials, interfaces, source kernel, and total power, so $R_{\mathrm{spread}}$ is a clean, non-negative residual. The temperature field is accumulated on one-dimensional and two-dimensional spatial bins, and every resistance is computed as
\begin{equation}
R=\frac{\Delta T}{P},
\end{equation}
with $\Delta T$ the source-to-sink temperature difference and $P$ the total injected power.

Because the atomistic and mesoscopic scales do not coincide, each method is normalized independently. Resistances are reported in dimensionless form,
\begin{equation}
R^{*}=\frac{R}{R_{\mathrm{ref}}},
\end{equation}
where $R_{\mathrm{ref}}$ is the one-dimensional resistance of that method's own bare-GaN reference. The two methods are therefore compared only on dimensionless trends and never on one another's absolute values. For the MC, the source heat flux is fixed at $q_{\mathrm{flux}}=10^{9}$~W~m$^{-2}$; the narrow and full-width sources inject the same total power $P_n=q_{\mathrm{flux}}w_g$, with the full-width flux density reduced by the width ratio.

\subsection{Equilibrium and LO non-equilibrium heat sources}

The equilibrium and non-equilibrium limits are realized by two atomistic heat sources. In the equilibrium reference, a $10$~\AA\ thick source slab is coupled to an enhanced-exchange algorithm that locally rethermalizes the atoms and injects a fixed power $P_{\mathrm{in}}=1.53$~eV~ps$^{-1}$. A $10$~\AA\ slab at the bottom is held at the sink temperature by a Langevin thermostat, while the remaining atoms evolve in the microcanonical ensemble.

In the non-equilibrium source, the Ga and N sublattices in the source region are driven in antiphase to reproduce the counter-propagating cation--anion motion of the zone-center longitudinal-optical mode. Each atom carries a frozen, uncorrelated random phase $\phi_i$, and the cross-plane forces are
\begin{equation}
F_{i,\mathrm{Ga}}=A\sin(\omega t+\phi_i),\qquad
F_{i,\mathrm{N}}=-A\sin(\omega t+\phi_i)\,\frac{N_{\mathrm{Ga}}}{N_{\mathrm{N}}},
\end{equation}
with amplitude $A=0.02$~eV~\AA$^{-1}$ and pump frequency $f=\omega/2\pi=25$~THz. The amplitude ratio $N_{\mathrm{Ga}}/N_{\mathrm{N}}$, where $N_{\mathrm{Ga}}$ and $N_{\mathrm{N}}$ are the numbers of Ga and N atoms in the source, enforces zero net force on the source, and the random phases over-populate LO phonons without launching a coherent wave. The injected power is an emergent quantity, the rate of work of the driving force,
\begin{equation}
P=\sum_i\mathbf{F}_i\cdot\mathbf{v}_i,
\end{equation}
averaged over complete pump cycles, where $\mathbf{v}_i$ is the velocity of atom $i$. Sampling at every time step prevents the $2\omega$ component of the 25~THz drive from aliasing into a spurious dc term.

\subsection{Gray and full-band phonon Monte Carlo}

The MC solver is a direct, event-driven solution of the steady-state phonon Boltzmann transport equation (BTE) along a ray direction $\mathbf{s}$,
\begin{equation}
\mathbf{s}\cdot\nabla f_\lambda=\frac{f_\lambda^{\mathrm{eq}}(T)-f_\lambda}{\tau_\lambda},
\end{equation}
where $\lambda=(\mathbf{q},p)$ labels a mode, $f_\lambda$ is the phonon distribution function, $f_\lambda^{\mathrm{eq}}(T)$ is its equilibrium value at temperature $T$, and $\tau_\lambda$ is the mode relaxation time. For the small source--sink difference ($10$~K) the deviation is linearized as $\hbar\omega f^d\approx C_\lambda\Delta T$, with per-mode heat capacity $C_\lambda=k_B x^2e^x/(e^x-1)^2$ and $x=\hbar\omega/k_BT$. With the reference at the sink, the sink deviation is zero and only the source injects energy. The body source emits isotropically over $4\pi$ with the deviation kernel
\begin{equation}
w_{\mathrm{emit}}\propto g(\omega)\,\hbar\omega\,\max\!\left[0,n_{\mathrm{BE}}(T_{\mathrm{emit}})-n_{\mathrm{BE}}(T_{\mathrm{amb}})\right],
\end{equation}
which contains no group velocity and reduces to the heat-capacity shape $g(\omega)C_\lambda\Delta T$ in the small-$\Delta T$ limit. Here $g(\omega)$ is the phonon density of states, $\hbar\omega$ is the phonon energy, $n_{\mathrm{BE}}$ is the Bose--Einstein occupation, and $T_{\mathrm{emit}}$ and $T_{\mathrm{amb}}$ are the emission and ambient temperatures. Each packet then advances by an exponentially distributed free path $\ell=-\Lambda_\lambda\ln R$, with $\Lambda_\lambda=v_g\tau$, and the nearest of bulk scattering, sidewall, or interface events is selected without a time step.

Such event-driven phonon MC has been applied across the ballistic-diffusive regime: it captures ballistic-diffusive conduction in silicon nanofilms \cite{47} and phonon wave propagation \cite{48}, while a full-band implementation has been compared with other simulation techniques for GaN HEMTs \cite{25}.

A gray model assigns one frequency-averaged mean free path to each material and is paired with the equilibrium molecular dynamics (MD). A full-band model draws dispersions, group velocities, densities of states, and relaxation times from a first-principles almaBTE workflow \cite{49}, and is required to distinguish the three structures. Each packet carries a bank flag identifying it as optical or acoustic. At a bulk event the post-scattering mode is drawn from the event-rate kernel $w_{\mathrm{scat}}\propto g(\omega)C_\lambda/\tau$, which is normalized separately over the acoustic and optical classes to give an acoustic and an optical landing spectrum. An optical packet down-converts with probability $p_{\mathrm{down}}=1/N_{\mathrm{op2ac}}$: on this event it is redrawn from the acoustic spectrum and flagged acoustic, and it stays acoustic thereafter because no acoustic-to-optical up-conversion is included; otherwise it is redrawn within the optical bank and remains optical. The number of optical scatterings before down-conversion is therefore a geometric variable with mean $N_{\mathrm{op2ac}}$, with $N_{\mathrm{op2ac}}=1$ recovering instantaneous single-spectrum re-thermalization, and the equivalent relaxation time is $\tau_{\mathrm{op2ac}}\simeq N_{\mathrm{op2ac}}\langle\tau_{\mathrm{op}}\rangle$. Each bulk event also re-emits the direction isotropically over $4\pi$ and resamples the free path from the new mode's mean free path. The temperature is obtained with the track-length estimator,
\begin{equation}
T(\mathrm{cell})=\frac{e_b\,t_{\mathrm{res}}}{V_{\mathrm{cell}}\,C_{\mathrm{vol}}},
\end{equation}
where $e_b$ is the energy carried by each packet, $t_{\mathrm{res}}$ is the accumulated physical path length divided by the group velocity, $V_{\mathrm{cell}}$ is the volume of the cell, and $C_{\mathrm{vol}}$ is the volumetric heat capacity of the material occupying that cell; this estimator is unbiased for any Knudsen number. Interfaces are elastic and frequency preserving, and are bracketed by a direction-aware, per-frequency diffuse-mismatch model and a specular acoustic-mismatch model based on Snell refraction and acoustic impedance \cite{50}. Non-equilibrium is introduced by selectively exciting only optical modes above a cutoff frequency, analogous to the LO-pumped MD. The direction-resolved diffuse-mismatch and specular acoustic-mismatch interface models are detailed in \textcolor{blue}{Supplementary Note S2}.

\subsection{Modal decomposition and stress analysis}

The partitioning of energy among modes is quantified by two complementary decompositions. The spectral energy density~\cite{51} projects the atomic velocities onto a wave vector,
\begin{equation}
V_\alpha(\mathbf{k},t)=\sum_i\sqrt{m_i}\,v_{i\alpha}(t)\,e^{i\mathbf{k}\cdot\mathbf{r}_i^0};
\end{equation}
where $\alpha$ denotes a Cartesian component, $m_i$ and $v_{i\alpha}$ are the mass and $\alpha$-component of velocity of atom $i$, and $\mathbf{r}_i^0$ is its equilibrium position. Fitting each mode's frequency peak to a Lorentzian gives the lifetime $\tau$ from the linewidth, and combining it with the harmonic group velocity gives the mean free path $\Lambda=v_g\tau$ and the cumulative thermal conductivity versus $\Lambda$. In the time-domain direct decomposition, the per-atom velocity time series is Fourier transformed and split into acoustic and optical bands, and the band mean-square velocity is converted to a band temperature through the equipartition relation,
\begin{equation}
T_{\mathrm{band}}=\frac{M\langle v^2\rangle_{\mathrm{band}}}{k_B},
\end{equation}
where $M$ is the atomic mass and $k_B$ is the Boltzmann constant. This yields separate acoustic and optical temperatures along the spreading path. The vibrational density of states, obtained from the Fourier transform of the velocity autocorrelation, is used to verify the band assignments.

The structural and elastic response is obtained from the per-atom virial stress,
\begin{equation}
\sigma_{\alpha\beta}=\frac{1}{V}\left[\sum_i m_i v_{i\alpha}v_{i\beta}+\sum_i r_{i\alpha}F_{i\beta}\right],
\end{equation}
where $\alpha$ and $\beta$ denote Cartesian components, $V$ is the volume, $m_i$ is the mass of atom $i$, $v_{i\alpha}$ and $v_{i\beta}$ are its velocity components, $r_{i\alpha}$ is its position component, and $F_{i\beta}$ is the force component acting on it. The stress is accumulated on the same spatial bins as the temperature field to map the stress around the source and interfaces. The elastic tensor and interface strain are obtained independently from finite-displacement force constants and from the relaxed atomic coordinates, providing a cross-check of the virial stress. Numerical settings, convergence tests, and full parameter files are given in \textcolor{blue}{Supplementary Notes S3 and S4}.

\section{Results and discussion}

\subsection{Equilibrium spreading: AlN shunt and AlGaN trap}

\subsubsection{Spreading dominates near-junction heat dissipation}

Under an equilibrium source, the near-junction thermal resistance is governed by lateral spreading rather than by one-dimensional conduction, and the AlN space layer suppresses this spreading component selectively while leaving the cross-plane resistance essentially unchanged.

For the GaN reference, the narrow-source spreading resistance is $R_{\mathrm{spread}}^{*}=3.28$, more than three times the one-dimensional resistance ($R_{\mathrm{1D}}^{*}=1.00$) and accounting for approximately 76\% of the total (\figref[a]{fig2}). The thermal resistance of a nanoscale hotspot is therefore incurred in the spatial constriction rather than along the bulk path. Introducing the AlGaN barrier raises $R_{\mathrm{1D}}^{*}$ to 4.63 and concurrently $R_{\mathrm{spread}}^{*}$ to 4.37, a 33\% increase relative to bare GaN. The one-dimensional increase is distributed over the GaN film rather than localized at the top interface; details are given in \textcolor{blue}{Supplementary Note S5}.

Inserting the AlN layer opens a missing lateral route. $R_{\mathrm{spread}}^{*}$ falls to 2.72, a 38\% reduction relative to the AlGaN/GaN structure, whereas $R_{\mathrm{1D}}^{*}$ remains essentially unchanged (4.63 to 5.10, +10\%). The AlN layer does not improve vertical conduction through the stack; it provides a high-thermal-conductivity lateral plane directly above the source that spreads upward-going energy before it is returned to GaN. The spreading resistance of the AlN-containing structure, 2.72, is approximately 17\% below that of the GaN film, 3.28. The benefit is therefore geometric and spectral rather than scalar.

\begin{figure}[htbp]
\centering
\includegraphics[width=\textwidth]{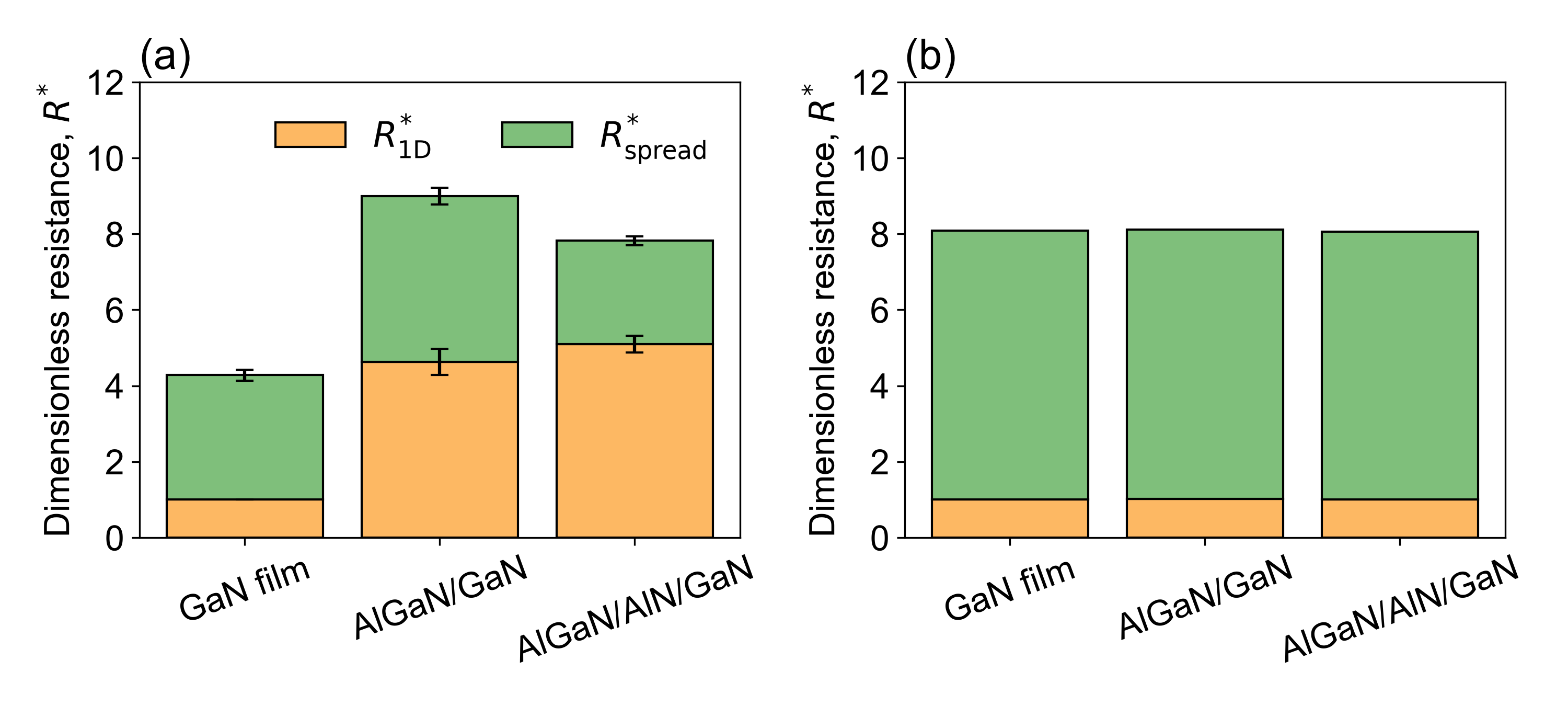}
\caption{Equilibrium spreading resistance of the three structures from MD and gray Monte Carlo. Stacked bars decompose the total resistance into one-dimensional, $R_{\mathrm{1D}}$, and spreading, $R_{\mathrm{spread}}$, components, normalized by the one-dimensional resistance of each method's own GaN reference. (a) MD: the GaN reference is dominated by spreading (76\%); the AlGaN barrier raises both components, while inserting AlN lowers $R_{\mathrm{spread}}$ from 4.37 to 2.72 with negligible change in $R_{\mathrm{1D}}$; error bars give the standard error across four independent seeds. (b) Gray MC: a frequency-averaged scalar description makes the structures nearly indistinguishable, $R_{\mathrm{spread}}=7.09/7.10/7.05$.}
\label{fig2}
\end{figure}

\subsubsection{Interface admission controls access; thickness is secondary}

The selective benefit observed for a single AlN thickness does not grow monotonically with thickness. Sweeping AlN thickness and interfacial bonding, parameterized by the bond-coupling factor $\chi$ around unity, shows that spreading resistance is controlled by interfacial coupling, and that thickness acts only insofar as the interface admits heat into the layer.

For the thickest layer, strengthening the interface from $\chi=0.9$ to 1.1 lowers $R_{\mathrm{spread}}^{*}$ from 3.53 to 2.18, a 38\% reduction; for the thinnest layer the same change is only approximately 7\%, with an almost flat response (\figref[a,c]{fig3}). $R_{\mathrm{1D}}^{*}$ stays within a narrow band, 4.80--5.60, across the entire grid (\figref[b]{fig3}), confirming that both parameters act selectively on spreading rather than on cross-plane conduction. The thickness trend reverses: at a weak interface, $\chi=0.9$, thickening the layer raises $R_{\mathrm{spread}}^{*}$ from 2.88 to 3.53; at a strong interface, $\chi=1.1$, thickening lowers it from 2.68 to 2.18.

The reversal orders the two parameters. Interfacial coupling determines whether heat enters the space layer, while thickness determines how much lateral distance is available once it does; access must be established before thickness can help. Details of the coupled response are provided in \textcolor{blue}{Supplementary Note S6}.

\subsubsection{Position determines whether high thermal conductivity is useful}

The benefit of a high-thermal-conductivity layer depends critically on where it is placed. When AlN is inserted near the junction, between AlGaN and GaN, it reduces the spreading resistance from 4.45 to 2.83 relative to AlGaN/GaN (\figref[d]{fig3}) by opening a lateral route inside the heat-generating region. The position and thickness cases in this section form an independent simulation series: their baseline values (4.45 and 2.83) differ slightly from the four-seed averages used for the three-structure ranking in \figref{fig2} (4.37 and 2.72), but the differences fall within the seed-to-seed statistical uncertainty. A thick Si heat sink beneath GaN lowers $R_{\mathrm{spread}}^{*}$ from 4.45 to 2.60 and the total from 9.29 to 7.47, but it does so by presenting an isothermal back boundary that widens the region over which the flow can expand, rather than by opening a lateral route inside the heat-generating volume. Once a near-junction AlN layer is present, this back-boundary benefit is smaller and is offset by added one-dimensional resistance, leaving the total essentially unchanged. The far-side substrate and the near-junction space layer thus act through different mechanisms and their benefits do not stack. A separate control reinforces the point: using the same AlN as the entire bottom substrate (AlGaN/GaN/AlN), its higher bulk thermal conductivity (about 285 versus 150~W~m$^{-1}$~K$^{-1}$ for Si) does not lower the resistance, and the structure gives the largest total of the substrate-mounted set ($R_{\mathrm{tot}}^{*}\approx11.4$). At the ten-nanometre scale the cross-plane boundary scattering conditions, rather than the substrate bulk thermal conductivity, set the film resistance; this is an ideal-coherent, cross-potential comparison not used for material ranking (details in \textcolor{blue}{Supplementary Note S5}). For the spreading resistance itself, a high-thermal-conductivity material embedded in the heat-generating pathway is therefore more effective than the same material added at the far boundary, which acts through a different mechanism.

\begin{figure}[htbp]
\centering
\includegraphics[width=\textwidth]{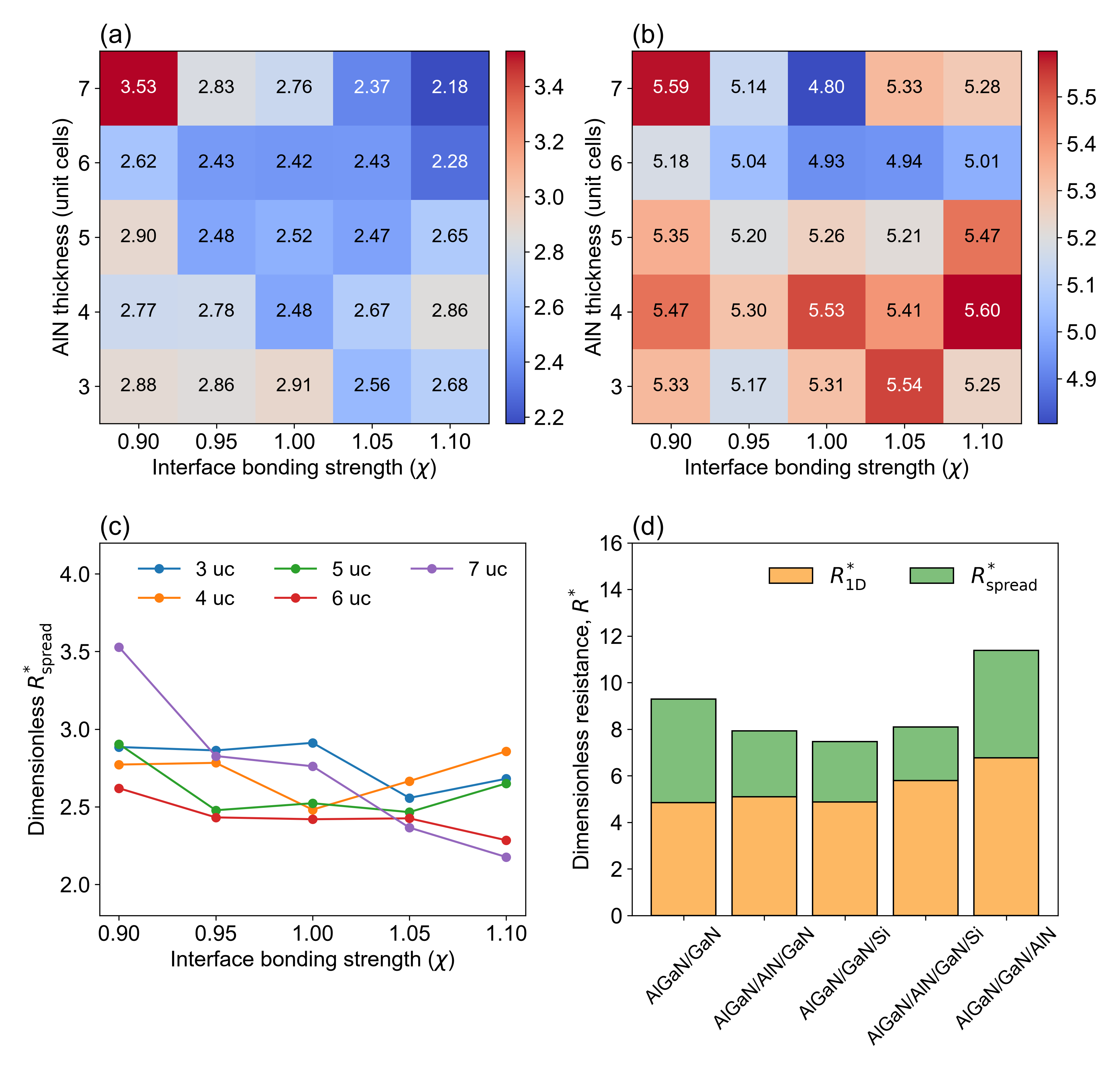}
\caption{Equilibrium spreading versus AlN thickness, interfacial bonding, and layer position. (a) Dimensionless $R_{\mathrm{spread}}^{*}$ and (b) $R_{\mathrm{1D}}^{*}$ over the thickness $\times$ bond-coupling grid. (c) $R_{\mathrm{spread}}^{*}$ versus $\chi$ for each thickness. (d) Dimensionless stacked resistance for the two free-standing structures and three alternative-position configurations. Interfacial coupling controls access to the layer; thickness and position matter only after access is established. AlN thickness is measured in unit cells (uc), i.e. the number of AlN crystal cells stacked along the growth direction. Here $\chi$ is the bond-coupling factor that scales the interfacial bond strength about the unmodified interface ($\chi=1$), and starred resistances are normalized by the one-dimensional Fourier-based reference $R_{\mathrm{1d0}}=t/(w\kappa_{\mathrm{GaN}})$.}
\label{fig3}
\end{figure}

\subsubsection{Spectral transmission, not scalar thermal conductivity, distinguishes structures}

To separate geometry and interface transmission from a scalar change in thermal conductivity, we revisit the three structures with a gray MC model that assigns each material a single, frequency-averaged mean free path. In this description the three structures are essentially indistinguishable ($R_{\mathrm{spread}}=7.09$, 7.10, and 7.05; \figref[b]{fig2}), with the spreading component dominant, approximately 88\%. The gray-model factorial analysis identifies the interface transmission and reflection coefficients, rather than the scalar mean free path or a scalar thermal conductivity, as the controlling variables: frequency averaging removes the spectrally resolved in-layer mean free paths that separate a shunt from a trap, while the interface coefficients retain their control.

On the AlN-containing geometry, the full-factorial main effects on $R_{\mathrm{tot}}^{*}$ (normalized by $R_{\mathrm{1D}}(s1)$, the same convention used in the SI figure) are dominated by the AlN/GaN interface transmission, $r=0.387$, followed by the AlGaN-side transmission, $q=0.240$. Interface specularity, $p=0.128$, and AlN thickness, $t=0.090$, are weaker, and all interaction terms are small. One-factor sweeps show $R_{\mathrm{spread}}$ falling markedly with $r$ and $q$ but remaining nearly flat against $p$ and $t$. Both methods locate the equilibrium spreading bottleneck at whether heat can cross the interface into a spreadable layer, rather than at bulk thermal conductivity.

\begin{figure}[htbp]
\centering
\includegraphics[width=\textwidth]{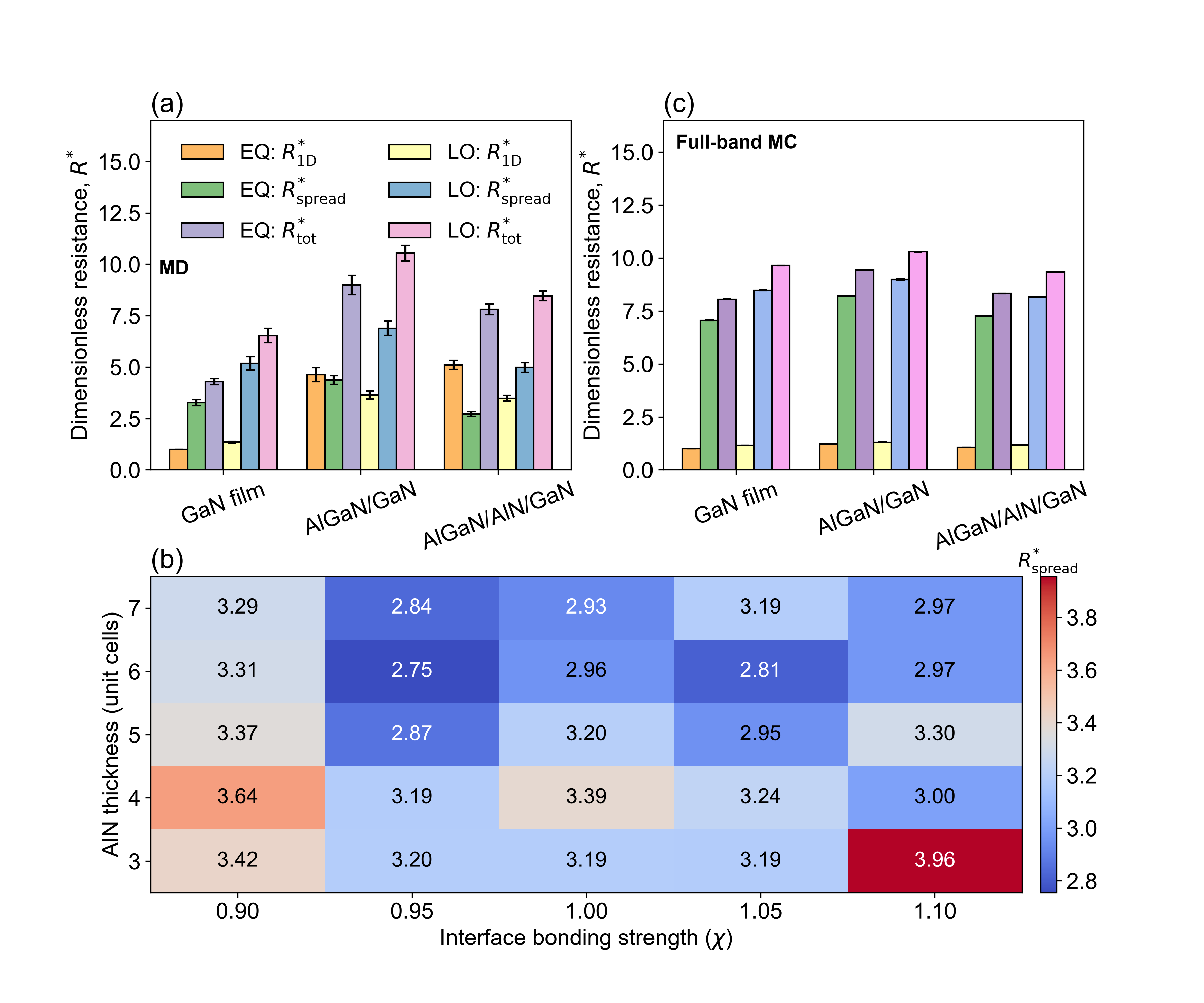}
\caption{Non-equilibrium spreading and weakened spatial control. (a) MD grouped bars for equilibrium and LO pumping, with error bars giving the standard error across four independent seeds. (b) Dimensionless $R_{\mathrm{spread}}^{*}$ over the LO thickness $\times$ bond-coupling grid. (c) Full-band MC grouped bars for equilibrium and LO pumping. LO pumping raises the spreading resistance in every structure, and the equilibrium spatial optimum weakens or becomes non-monotonic. Here $\chi$ is the bond-coupling factor that scales the interfacial bond strength about the unmodified interface ($\chi=1$), and starred resistances are normalized by the one-dimensional Fourier-based reference $R_{\mathrm{1d0}}=t/(w\kappa_{\mathrm{GaN}})$.}
\label{fig4}
\end{figure}

\subsection{Non-equilibrium spreading: the conversion gate becomes dominant}

\subsubsection{LO pumping increases spreading resistance}

Replacing the equilibrium source with a localized LO-phonon pump increases the spreading resistance for every structure and shifts the hotspot toward the interface (\figref[a]{fig4}). $R_{\mathrm{spread}}^{*}$ increases from 3.28 to 5.18 for GaN, +58\%; from 4.37 to 6.89 for AlGaN/GaN, +58\%; and from 2.72 to 4.98 for AlGaN/AlN/GaN, +83\%. The ordering is preserved, but the advantage of the AlN-containing structure is eroded relative to the equilibrium case.

The two-dimensional temperature fields show that the peak temperature is concentrated at the interface height where energy is injected into slow optical modes. The AlGaN/GaN structure makes the distinction between peak temperature and integrated resistance explicit: its peak is nearly identical to that of the AlN-containing structure, yet it carries the largest spreading resistance. Because spreading resistance is an integral of the temperature excess over the constriction rather than the value at the single hottest cell, a warm but unspread structure outranks a hotter one that spreads effectively. Details of the temperature-field morphology are given in \textcolor{blue}{Supplementary Note S7}.

This morphology is the spatial signature of a spectral bottleneck. Optical modes have low group velocities and cannot carry heat laterally, so the energy they hold cannot spread until it relaxes into acoustic modes. Under an equilibrium source this conversion is already effectively complete. The near-junction resistance therefore depends not only on geometry but also on the spectral state of the injected energy.

\subsubsection{Equilibrium spatial controls weaken or reverse}

Under LO heating the entire response surface shifts upward and becomes flatter ($R_{\mathrm{spread}}^{*}\approx2.75$--3.96 versus 2.18--3.53 under equilibrium; \figref[b]{fig4}), and the strong influence of interfacial coupling on thick AlN is largely lost. At $\chi=1.0$, increasing thickness reduces $R_{\mathrm{spread}}^{*}$ only from 3.19 to 2.93, $-8\%$. The increment map $\Delta R_{\mathrm{spread}}^{*}=R_{\mathrm{LO}}^{*}-R_{\mathrm{EQ}}^{*}$ shows that the increase is almost universal: $\Delta R_{\mathrm{spread}}^{*}>0$ in 24 of the 25 cells, the sole exception being the thick-AlN/weak-interface cell, and the increment peaks at the thin-AlN/strong-interface cell. The thin-AlN/strong-interface cell is favourable under equilibrium but becomes the global maximum under LO, whereas the equilibrium optimum, thick AlN at the strongest interface, is no longer recovered.

The reversal exposes the precondition for spatial-layer control. When energy already resides in mobile acoustic modes, interface transmission determines whether that energy can reach a spreadable layer. Under LO pumping, energy is trapped in slow optical modes and the rate-limiting step is optical-to-acoustic conversion within the source, which changing interface transmission does not accelerate. Geometry and interface control therefore operate only after spectral conversion has occurred.

\subsubsection{Spectral resolution is required}

The full-band, spectrally resolved MC independently reproduces the non-equilibrium increase: $R_{\mathrm{spread}}$ rises from 7.05 to 8.47 for the bare-GaN structure (s1), +20\%; from 8.19 to 8.96 for AlGaN/GaN (s2), +9\%; and from 7.27 to 8.15 for AlGaN/AlN/GaN (s3), +12\%, in qualitative agreement with the NEMD results (\figref[c]{fig4}).

The decisive comparison is with the gray model, which shows no non-equilibrium sensitivity, whereas the full-band model both distinguishes the structures and resolves their non-equilibrium increase. Once frequency-dependent group velocities, lifetimes, and separate optical and acoustic reservoirs are retained, structural and non-equilibrium effects appear together. To determine the controlling parameter and its tunable range, optical-to-acoustic conversion must itself be swept and incorporated into a non-equilibrium resistance model.

\subsection{Three-gate mechanism and predictive resistance network}

\subsubsection{Spectral basis: acoustic carriers and pumped optical modes}

\begin{figure}[htbp]
\centering
\includegraphics[width=\textwidth]{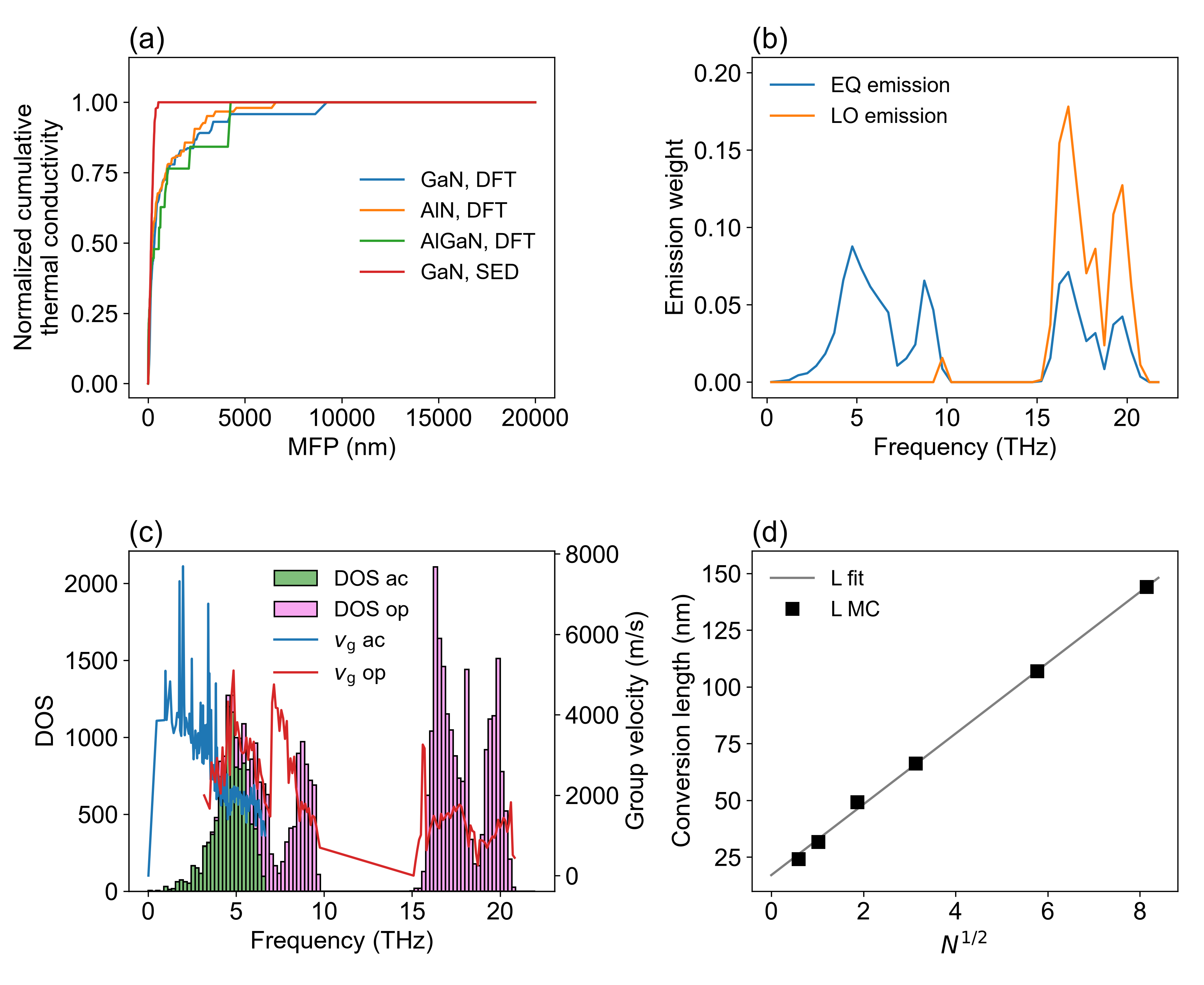}
\caption{Spectral basis of near-junction spreading and the conversion length. (a) Cumulative thermal conductivity versus MFP from DFT for GaN, AlN, and AlGaN, and from spectral-energy decomposition with the Tersoff potential for GaN, each normalized to its bulk value. (b) Source emission spectra under equilibrium and LO pumping. (c) GaN density of states and group velocity. (d) Conversion length $L$ versus $\sqrt{N/3}$, with the MC values and the random-walk fit.}
\label{fig5}
\end{figure}

\begin{figure}[htbp]
\centering
\includegraphics[width=\textwidth]{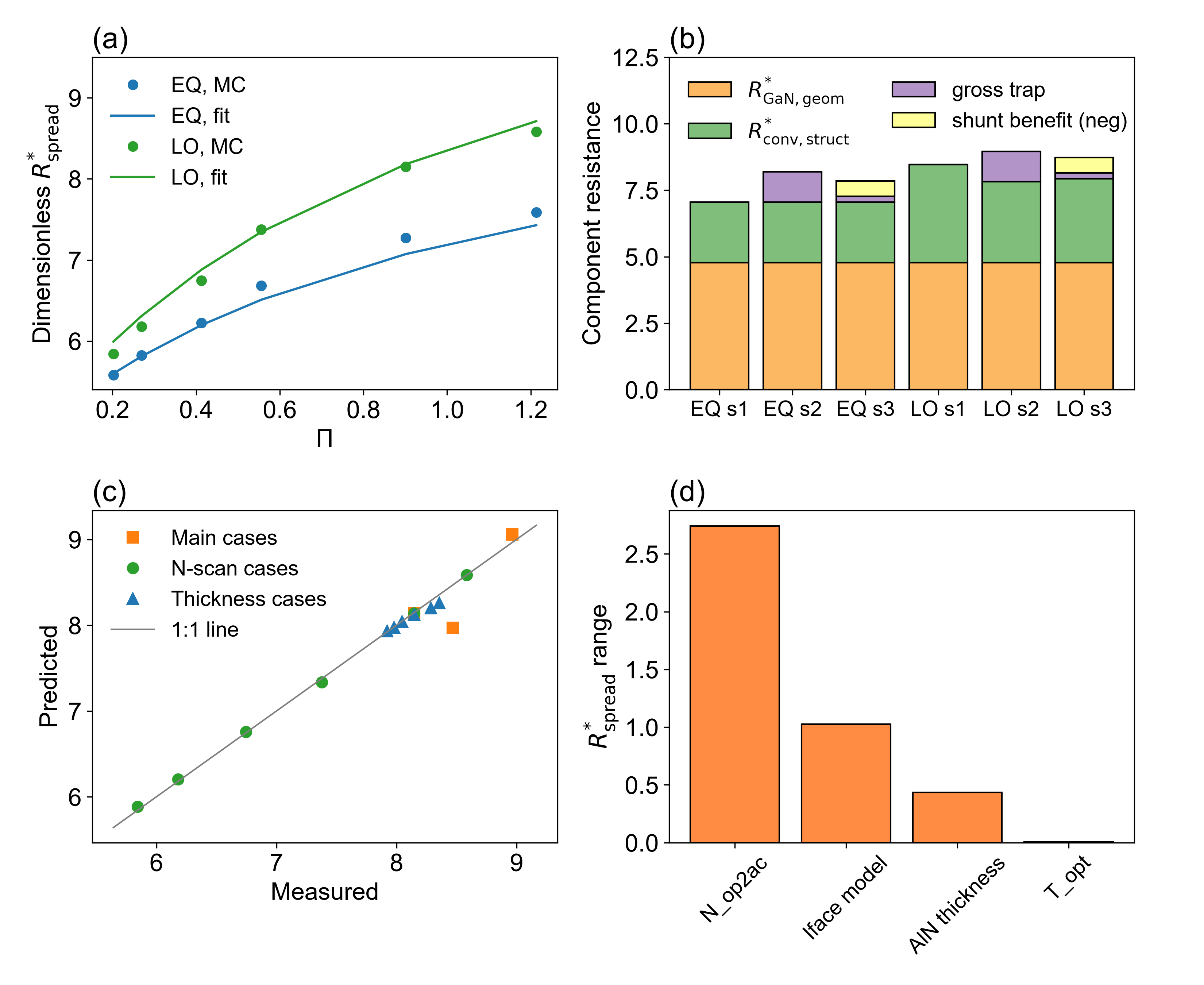}
\caption{Predictive resistance network, validation, and sensitivity. (a) Conversion increment $R_{\mathrm{conv}}^{*}$ versus the non-equilibrium conversion number $\Pi$, showing the saturating $\Pi/(1+\Pi)$ form. (b) Component decomposition for the three structures under equilibrium and LO heating. (c) Parity of measured versus predicted resistance. (d) One-factor-at-a-time sensitivity of $R_{\mathrm{spread}}^{*}$. The network closes the three-gate mechanism onto a small set of dimensionless groups. Here s1, s2, and s3 denote the bare GaN, AlGaN/GaN, and AlGaN/AlN/GaN structures, respectively.}
\label{fig6}
\end{figure}

The resistance number alone cannot be designed. To turn the equilibrium--non-equilibrium contrast into a predictive model, we resolve the path that a source phonon must traverse before its energy reaches the sink. Three gates must open in sequence, and closing any one cannot be compensated by the others (\textcolor{blue}{Figs.~\ref{fig5}--\ref{fig6}}). Gate~0 is a spectral conversion: energy in optical modes must relax into acoustic modes before it can be transported. Gate~1 is interface admission: source phonons must be spectrally transmitted across the interface into the spreadable layer. Gate~2 is in-layer diversion: once admitted, the energy must be carried laterally, which requires the layer to overlap the spreading region and provide enough lateral distance.

Spreading is carried by long-MFP, in-plane acoustic modes, whereas the hot-carrier-pumped modes are nearly stationary; the two reservoirs are connected only by optical-to-acoustic conversion. The density-functional-theory (DFT) cumulative thermal conductivity makes this separation explicit (\figref[a]{fig5}): half of the GaN thermal conductivity is carried by phonons with MFP above 324~nm and 90\% above approximately 3.2~$\mu$m. This 324~nm half-contribution cutoff agrees closely with the independent first-principles cumulative calculation of Ref.~\cite{25}, which locates the same half-contribution MFP at about 300~nm and shows that an effective-MFP model fitted to 1000~nm overestimates the controlling length; the two atomistic estimates differ by only about 8\% and jointly establish the sub-micron acoustic scale. AlN reaches 241~W~m$^{-1}$~K$^{-1}$ with a larger acoustic share, while the AlGaN alloy collapses to $\kappa\approx13.6$~W~m$^{-1}$~K$^{-1}$ with a heat-capacity-weighted MFP of only 4~nm. The Tersoff spectral-energy analysis gives the same cumulative GaN shape despite its smaller absolute value, so the conclusion is robust: spreading is the work of sub-micron to micron acoustic modes. The source instead injects a distinct population: the high-frequency optical modes at and above 9.55~THz carry only 2.2\% of the GaN thermal conductivity, with MFP $\approx$10~nm, relaxation time 7.7~ps, and group velocity 1266~m~s$^{-1}$ (\figref[b,c]{fig5}). The equilibrium source emits a broad near-thermal spectrum, with optical fraction $\eta_{\mathrm{op}}=0.673$, whereas the LO source emits entirely above 9.55~THz; energy in these modes is locked at the source until it converts downward (details in \textcolor{blue}{Supplementary Note S4}).

\subsubsection{Conversion bottleneck}

Under LO pumping the source enters a two-reservoir state (optical modes over-populated and acoustic modes under-populated), and the length over which this imbalance relaxes is the gate that ultimately limits non-equilibrium spreading.

The spectral temperatures (\textcolor{blue}{Supplementary Note S4}) expose both faces: in NEMD the source shows $\delta_{\mathrm{op}}=+11.4\pm0.5$~K and $\delta_{\mathrm{ac}}=-6.4\pm0.7$~K under LO pumping, versus a weak equilibrium split $+2.9$ and $-1.6$~K. The full-band MC reproduces this and, by sweeping the conversion parameter, shows how it is controlled: at slow conversion $N=100$ the optical temperature peaks inside the source and stays elevated at 80~nm depth, whereas at instantaneous conversion $N=1$ the optical excess peaks at only 0.62~K and decays within tens of nanometres.

The branch-resolved field shows this is a bottleneck of the specific pumped modes rather than a generic optical excess (\textcolor{blue}{Supplementary Note S6}). Under LO pumping only the high-frequency optical branches above 9.55~THz with approximately 10~nm MFP are over-populated ($+5$ to $+6\%$), while the acoustic and transport-active low-frequency optical branches with MFP around 48~nm are jointly under-heated ($-2$ to $-3\%$). The pump locks energy into precisely the nearly stationary reservoir of \figref[b]{fig5} and leaves the acoustic channels cold.

The spatial extent of this locked-in energy is set by a conversion length extracted from the thermal field by both methods (\figref[d]{fig5}): in NEMD the near-source optical excess decays exponentially with $L_{\mathrm{op}\to\mathrm{ac}}=2.37$~nm, while in the full-band MC the characteristic length into the GaN bulk grows from 24 to 144~nm as $N$ runs from 1 to 200 and collapses onto the random-walk form $L=L_0+\ell\sqrt{N/3}$, with a ballistic leg $L_0\approx17$~nm and a random-walk step $\ell\approx15.6$~nm. These are thermal-field decay lengths including conversion and diffusion, not pure particle-level survival lengths; the absolute values are not transferred between methods (the Tersoff potential, nanometre source, and fast intrinsic conversion give approximately 2.4~nm, while the device-scale DFT spectrum spans 17--144~nm). Only the $\sqrt{N}$ scaling, and hence the diffusive origin, is shared (particle-level statistics in \textcolor{blue}{Supplementary Note S8}).

The two-reservoir split we find is the same phenomenon now resolved quantitatively in independent experiments and coupled simulations. Near a 7.2~nm lateral (2.1~nm vertical) GaN hotspot, Raman thermometry through a 79-fold field-enhanced nanotip measures A$_{1}$(LO) and E$_{2}$(TO) temperature rises of 26.5 and 10.2~K at 5.42~mW, while the acoustic rise stays lower by nearly two orders of magnitude (about 0.3~K), consistent with roughly 94\% of the deposited energy entering LO modes \cite{39}; this optical-hot/acoustic-cold ordering matches our $\delta_{\mathrm{op}}=+11.4$ and $\delta_{\mathrm{ac}}=-6.4$~K. Coupled electron--phonon MC gives the same imbalance at device level: an AlGaN/GaN FinFET hotspot reaches 352~K, its dominant TO2 branch running 80\% above the lattice temperature and supplying 63\% of phonon emission \cite{40}, and under 2\% compressive strain the peak heat generation rises 6.2-fold and the hotspot temperature from 42 to 282~K (582~K under full self-heating) \cite{41}. At GaN/AlN superlattice interfaces the mode-temperature drop accounts for about 69\% of the total drop and is carried almost entirely by the acoustic and first-optical branches \cite{38}. These reported values are modal or hotspot-peak temperatures rather than normalized spreading-resistance increments, so their magnitudes are not equated with our 58--83\% (NEMD) and 9--20\% (MC) resistance rises; what transfers is the robust direction---selective optical excitation, cold acoustic channels, and an elevated non-equilibrium resistance.

\subsubsection{Strain and interface admission}

The interface admits heat through spectral overlap rather than a scalar normal conductance, and the buried AlN opens this channel through coherent misfit strain present only in the atomistic description. The equilibrium in-plane stress fields (\textcolor{blue}{Supplementary Note S6}) show why: a bare GaN film is nearly stress-free (in-layer peak $|\sigma_{xx}|\approx2.8$~GPa), an AlGaN cap introduces a coherent compressive band ($\approx$25~GPa), and inserting AlN extends and strengthens it across AlN into the adjacent GaN (peak $\approx$47~GPa); under LO pumping the band reorganizes at the source. This coherent strain shifts the interfacial force constants and reorganizes the interfacial phonon spectrum, whereas MC treats the interface as a spectral boundary condition with no strain. Yet the field is not confined to the interface: AlN adds only $+0.95$~GPa to the uniform GaN tensile background (biaxial strain $\Delta e=0.23\%$), over which the in-plane longitudinal group velocity shifts by only $-0.30\%$ and the bulk thermal conductivity responds by less than 1\%, two orders of magnitude below the 38\% reduction of $R_{\mathrm{spread}}$. The AlN space layer therefore acts through a new high-thermal-conductivity channel and its interfacial spectrum, not through a strain-induced change of GaN bulk.

This is why interface coupling acts almost entirely on spreading rather than on one-dimensional resistance: a scalar Kapitza conductance would change both equally. The coupling instead controls lateral admission---whether a source phonon can enter high-thermal-conductivity AlN and then be diverted in-plane---so the upper boundary condition must retain spectral transmission rather than collapse to a single normal conductance.

\subsubsection{In-layer lateral diversion}

Once admitted, energy is diverted laterally, and the strength of this gate is geometric---whether the layer overlaps the spreading region and offers enough lateral distance---rather than set by material thermal conductivity alone. The layer-resolved NEMD lateral flux (\textcolor{blue}{Supplementary Note S6}) localizes where spreading happens: under LO pumping the AlN layer just outside the source carries $0.53\pm0.14$~eV~ps$^{-1}$ outward over the near edge, 2.4 times the GaN source slab ($0.22\pm0.07$~eV~ps$^{-1}$); further away the GaN flux decays to 0.05--0.07~eV~ps$^{-1}$ and becomes statistically zero, whereas the AlN flux stays at 0.24--0.30~eV~ps$^{-1}$, 4--5 times the GaN level. This in-plane AlN flux, set at the source edge yet sustained far along the layer, is the signature of a spreader with high in-plane thermal conductivity; the equilibrium panel is less conclusive.

The device-scale MC shows the same edge-localized diversion and its geometric limit (\textcolor{blue}{Supplementary Note S6}). The AlN lateral flux peaks at the source edge and returns to zero within a few micrometres; increasing AlN thickness from 1 to 20~nm raises the integrated flux about sixfold, yet for a fixed ideal interface $R_{\mathrm{spread}}$ falls only approximately 4\% in equilibrium and approximately 5\% under LO pumping, because the layer is far thinner than the source and cannot occupy enough of the spreading cone. In the atomistic model, where AlN is comparable to the source width, the diversion is large and the thickness--coupling response non-monotonic. The two scales are complementary---MC isolates the clean geometric benefit, NEMD adds the real interface that switches it on---and position remains the other geometric condition, that the high-thermal-conductivity layer overlap the production region.

\subsubsection{Predictive thermal-resistance network}

The three gates become a design tool when they map onto a small set of dimensionless groups and assemble into a closed network with known limits. Starting from the single-layer ballistic--diffusive model (a GaN film with an equilibrium boundary source and adiabatic upper surface), we make two extensions: replace the adiabatic top by a partially transmitting cap unit (the interface-admission gate plus the cap layer), and replace the equilibrium source by a non-equilibrium optical pump that places the conversion gate ahead of all spatial transport.

All resistances are normalized by the one-dimensional Fourier-based reference $R_{\mathrm{1d0}}=t/(w\kappa_{\mathrm{GaN}})$, where $t$ is GaN thickness, $w$ the period width, and $\kappa_{\mathrm{GaN}}$ the bulk thermal conductivity. For a source of width $w_g$ on an adiabatic top, isothermal bottom, and periodic sides, the gray ballistic--diffusive result factors as~\cite{21}
$R_{\mathrm{GaN}}/R_{\mathrm{1d0}}=(R_{\mathrm{F}}(\varepsilon,t/w)/R_{\mathrm{1d0}})(1+\tfrac{2}{3}Kn_t)(1+A_w Kn_w)$,
where $\varepsilon=w_g/w$, $Kn_t=\Lambda/t$, $Kn_w=\Lambda/w_g$, $R_{\mathrm{F}}$ is the Fourier-based spreading series, and $2/3$ is the single-boundary Marshak jump coefficient; retaining the full spectrum replaces the MFP corrections by material-independent functions (\textcolor{blue}{Supplementary Note S9}).

A non-equilibrium source adds a resistance controlled by phonons rather than geometry. The source emits a fraction $\eta_{\mathrm{op}}$ of its power into slow optical modes---an emission-spectrum input (0.673 for the near-thermal equilibrium source, 1.000 for the LO pump), not an on/off switch. In steady state the two reservoirs obey $0=\eta_{\mathrm{op}}\dot q-G_{\mathrm{conv}}(T_{\mathrm{op}}-T_{\mathrm{ac}})$, with $G_{\mathrm{conv}}=C_{\mathrm{op}}/\tau_{\mathrm{conv}}$. Because the pumped optical modes above 9.55~THz carry only 2.2\% of the thermal conductivity, this energy must convert before it can spread; it cannot bypass the gate. The controlling dimensionless group is the non-equilibrium conversion number,
\begin{equation}
\Pi=\frac{\tau_{\mathrm{conv}}}{\tau_{\mathrm{ac}}}
=\frac{L_{\mathrm{op}\to\mathrm{ac}}}{\Lambda_{\mathrm{ac}}},
\end{equation}
where $\tau_{\mathrm{conv}}$ is the optical-to-acoustic conversion relaxation time and $\tau_{\mathrm{ac}}$ is the acoustic transport time, with $L_{\mathrm{op}\to\mathrm{ac}}$ the decay length of the optical over-population field and $\Lambda_{\mathrm{ac}}$ the geometry-confined, in-plane MFP of the heat-carrying acoustic modes. The two-reservoir balance then gives, as seen in \figref[a]{fig6}, the saturating harmonic form $R_{\mathrm{conv}}^{*}=\eta_{\mathrm{op}}C_c\Pi/(1+\Pi)$, where $C_c$ is a single order-unity coefficient set by the two-temperature solution. Because the equilibrium source already carries a finite optical fraction and conversion slip, the physically reported conversion resistance is the excess over the equilibrium baseline rather than the absolute value,
\begin{equation}
R_{\mathrm{conv}}^{*}
=C_c\left[\eta_{\mathrm{op}}\frac{\Pi}{1+\Pi}
-\eta_{\mathrm{op}}^{\mathrm{eq}}\frac{\Pi_{\mathrm{eq}}}{1+\Pi_{\mathrm{eq}}}\right],
\end{equation}
which vanishes identically at equilibrium ($\eta_{\mathrm{op}}=\eta_{\mathrm{op}}^{\mathrm{eq}}$, $\Pi=\Pi_{\mathrm{eq}}$), making the equilibrium spatial network the reference and isolating the conversion cost imposed by the LO pump.

The measured lengths place the conversion gate across the observed crossover. In the full-band MC, $L_{\mathrm{op}\to\mathrm{ac}}$ runs from 24 to 144~nm as $N$ goes from 1 to 200, so $\Pi$ rises from approximately 0.20 to 1.21 for $\Lambda_{\mathrm{ac}}=118.7$~nm: fast conversion ($N\lesssim30$, $\Pi<0.6$) keeps the gate nearly transparent, $N=100$ is critical ($\Pi\approx0.90$), and $N=200$ closes it ($\Pi>1$). In NEMD, where the nanometric film confines the acoustic MFP to roughly its thickness, $L_{\mathrm{op}\to\mathrm{ac}}=2.37$~nm against $\Lambda_{\mathrm{ac}}\approx10.4$~nm gives $\Pi\approx0.23$, precisely the fast-conversion end. The two engines share only the shape of the $\Pi$ dependence, not the absolute lengths.

The layer above the source is neither an adiabatic wall nor a one-way shunt, but an interface-plus-layer unit that returns part of the admitted energy to the hotspot and diverts the rest. The interface gate is parameterized directly by spectral energy transmission $T_{\mathrm{if}}$ ($0\leq T_{\mathrm{if}}\leq1$), supplied mode by mode by the direction-resolved diffuse-mismatch model, with the acoustic-mismatch model as the specular bound. For lateral spreading only the in-plane component can feed a shunt, so the transmission is weighted by the post-transmission angular distribution, $T_{\mathrm{if}}^{\mathrm{lat}}=T_{\mathrm{if}}\langle g(\theta)\rangle$. A diffuse-mismatch interface re-emits into a Lambertian field with large in-plane weight, whereas a specular acoustic-mismatch interface conserves tangential momentum and sends near-normal phonons straight through; with the adiabatic top allowing no vertical exit, that normal component cannot drain and is retained. A specular interface that helps a vertical heat sink can therefore hinder a lateral shunt.

Whether that admitted energy is then removed is decided within the cap. The in-layer conductance is set by the cap Knudsen number $Kn_{\mathrm{cap}}=\Lambda_{\mathrm{cap}}/t_{\mathrm{cap}}$, the aspect ratio $\varepsilon_{\mathrm{cap}}=t_{\mathrm{cap}}/w_g$, a constrained-conduction shape factor, and a source-overlap function (full expression in \textcolor{blue}{Supplementary Note S9}), giving the net shunt conductance $G_{\mathrm{shunt}}^{*}=T_{\mathrm{if}}^{\mathrm{lat}}G_{\mathrm{lat}}^{*}$.

Energy that enters the cap but is not laterally removed has nowhere to go under the adiabatic top and must return. The cap diversion efficiency $\eta_{\mathrm{lat}}\in[0,1]$ is the fraction of admitted energy carried away laterally; the complementary fraction $1-\eta_{\mathrm{lat}}$ is retained and re-injected at the hotspot. This angle-randomized return presents the GaN top boundary with an effective diffuse reflectivity, $r_{\mathrm{eff}}=T_{\mathrm{if}}(1-\eta_{\mathrm{lat}})$, which amplifies the boundary temperature jump in the slip form of the interfacial BTE,
\begin{equation}
R_{\mathrm{trap}}^{*}=C_t r_{\mathrm{eff}}\frac{1+r_{\mathrm{eff}}}{1-r_{\mathrm{eff}}},
\end{equation}
with $C_t$ an order-unity coefficient. The cap therefore contributes two opposing terms: a shunt conductance that lowers resistance and a retained-flow boundary slip that raises it.

The diversion efficiency follows from the residence probability of body scattering before the interface return,
\begin{equation}
\eta_{\mathrm{lat}}=\frac{2}{3}\frac{\Lambda_{\mathrm{cap}}T_{\mathrm{if}}}{4t+\Lambda_{\mathrm{cap}}T_{\mathrm{if}}},
\end{equation}
where $\Lambda_{\mathrm{cap}}$ is the phonon MFP in the cap and $t$ is the cap thickness. This separates the regimes by construction: the long-MFP AlN cap ($\Lambda_{\mathrm{cap}}=92.3$~nm) gives $\eta_{\mathrm{lat}}\approx0.43$ (a shunt), while the short-MFP AlGaN cap ($\Lambda_{\mathrm{cap}}=7.5$~nm) gives $\eta_{\mathrm{lat}}\approx0.03$ (a trap). The kinetic boundary condition gives the slip coefficient $C_t=0.93$; the full spectral BTE supplies only the residual thickness curvature, left to \textcolor{blue}{Supplementary Note S9}.

The three gates assemble into one network, whose component decomposition is illustrated in \figref[b]{fig6}. Using $A\parallel B=AB/(A+B)$ for parallel resistances, the non-equilibrium near-junction resistance is
\begin{equation}
R_{\mathrm{near}}^{*}
=
R_{\mathrm{conv}}^{*}(\Pi,\eta_{\mathrm{op}})
+
\left[
\left(R_{\mathrm{GaN}}^{*}+R_{\mathrm{trap}}^{*}\right)
\parallel
R_{\mathrm{shunt}}^{*}
\right],
\end{equation}
where $R_{\mathrm{GaN}}^{*}$ is the near-junction resistance of the GaN layer and the shunt resistance is $R_{\mathrm{shunt}}^{*}=1/(T_{\mathrm{if}}^{\mathrm{lat}}G_{\mathrm{lat}}^{*})$. The network contains six independent dimensionless groups: $\Pi$, $\eta_{\mathrm{op}}$, $T_{\mathrm{if}}$, $Kn_{\mathrm{cap}}$, $\varepsilon_{\mathrm{cap}}$, and $z_{\mathrm{if}}/w$; angle-resolved transmission and cap diversion efficiency are derived combinations rather than new fitted parameters. The network is closed by its limiting cases. Instantaneous conversion ($\Pi\to0$, $N=1$) gives $R_{\mathrm{conv}}^{*}\to0$ for both source states even though $\eta_{\mathrm{op}}$ keeps its fixed spectrum (0.673 equilibrium, 1.000 LO). At the production point $N=100$, $\Pi\approx0.90$, the baseline-subtracted term vanishes at equilibrium by construction and is finite under LO pumping ($R_{\mathrm{conv}}^{*}\approx0.87$ for the AlN structure, up to $\sim1.4$ for bare GaN). A bare GaN film ($T_{\mathrm{if}}=0$) reduces to the single-layer baseline, as both shunt and trap vanish. A low-thermal-conductivity AlGaN cap ($\eta_{\mathrm{lat}}\approx0$) can only add resistance, because the shunt vanishes while retained-flow slip stays finite; a high-thermal-conductivity AlN cap with large $\eta_{\mathrm{lat}}$ has weak slip and a strong shunt, pulling resistance back toward the bare-film value. A closed interface ($T_{\mathrm{if}}\to0$) removes both cap effects regardless of material.

The coefficients are not free fitting numbers but carry identifiable kinetic and geometric origins. Across the entire network only two order-unity coefficients, $C_c$ and $C_t$, are fixed against the simulation rather than calibrated separately for each case, and the phenomenological conversion parameter $N_{\mathrm{op2ac}}$ is a physically defined engineering input that is swept over a range rather than fitted. The conversion length is a random walk: $N$ is the number of optical bulk scatterings before down-conversion ($p_{\mathrm{down}}=1/N$), each randomizing the direction, so the cross-plane displacement grows as the free-flight step times $\sqrt{N/3}$. The same step computed directly from the scattering-weighted spectrum is $\Lambda_{\mathrm{op}}^{\mathrm{scat}}\approx14.1$~nm, reproducing the fitted $\ell=15.6$~nm and $L_0=17.0$~nm within 9.6--17.1\%; a purely spectral, un-fitted prediction of the six conversion lengths has only 12\% error. The conversion amplitude likewise separates into a first-principles heat-capacity ratio and a geometric factor,
$C_c=(C_{\mathrm{op}}/C_{\mathrm{ac}})g(\varepsilon)$, with $C_{\mathrm{op}}/C_{\mathrm{ac}}=2.03$ and $g(\varepsilon)\approx\ln(4w/\pi w_g)=2.54$, giving $C_c\approx5.2$ versus the fitted 5.58. The cap coefficient $C_t=0.93$ and $\eta_{\mathrm{lat}}$ are likewise evaluated from the kinetic boundary condition and the dimensionless groups rather than fitted case by case. Thus the functional forms follow from kinetic saturation and random walks, the optical fraction and interface transmission are first-principles inputs, and the full spectral BTE supplies only the residual thickness curvature rather than a device-level fitted parameter.

With $C_c=5.582$ and the independently evaluated $C_t=0.93$ and $\eta_{\mathrm{lat}}$, the network predicts, as illustrated in \figref[c]{fig6}, the six full-band Monte Carlo cases (three structures, equilibrium and LO pumping) to within $1.6\%$ for five of them; the remaining case, bare GaN under LO pumping, is within $6.5\%$, the excess being the second-order Gate-0 $\times$ cap interaction. The earlier fitted calibration and full leave-one-out/holdout statistics are in \textcolor{blue}{Supplementary Note S9}; the model is calibrated against the full-band MC, while MD values are compared at the trend level. It converts directly into design rules: accelerate optical-to-acoustic conversion to keep $\Pi<1$; engineer interfaces for large laterally directed spectral transmission; make the cap itself highly conductive; and treat thickness and position as secondary, since they act only after the preceding gates and require the layer to overlap the source. The one-factor-at-a-time ranking in \figref[d]{fig6} gives the same ordering, with interface transmission and conversion dominating over thickness.

It is noted here that this quantitative agreement holds at the device scale, where both descriptions treat the lattice vibrations as incoherent particles. Beyond this quantitative agreement, the two methods differ in whether they retain the wave character of the lattice vibrations, which sets the length scale over which their agreement should be expected. The atomistic AlN space layer is only 1.56~nm thick, comparable to the wavelengths of the heat-carrying phonons; because molecular dynamics evolves the atomic displacements directly, it resolves the interference, interface-mode hybridization and phonon confinement arising on that scale. The phonon Monte Carlo solution treats each phonon as an independent ray packet under the Boltzmann equation, carrying energy and direction but no phase, so coherent interference is absent. The two calculations also differ in domain size, interface treatment and spectral resolution, so their comparison is made at the trend level; the loss of structural contrast in the gray model, in particular, follows from frequency averaging. On the points that define the spreading mechanism—the dominance of spreading over one-dimensional conduction, the control by interface admission, and the LO-induced increase with its weakened spatial gain—the atomistic and particle descriptions agree qualitatively, showing that the incoherent Boltzmann picture captures the leading transport at the scales examined here. The coherent, wave-resolved regime, which enters at the 1.56~nm layer scale set by the phonon wavelengths, forms the lower length-scale boundary of this particle model.

\section{Conclusions}

With the AlN space layer treated as a tunable design element rather than a passive material, we have directly tuned the near-junction heat spreading across the GaN, AlGaN/GaN, and AlGaN/AlN/GaN structures under equilibrium and localized LO-pumped heat sources, using non-equilibrium molecular dynamics and full-band spectral phonon Monte Carlo as the examining tools rather than as the focus. The results show that near-junction heat spreading is governed by coupled spatial and spectral gates that a scalar thermal conductivity cannot capture, so that the benefit of a high-thermal-conductivity layer depends on the state of the source. These findings lead to four conclusions.

(1) Under equilibrium heating, the AlN space layer tunes spreading through interfacial admission rather than thickness: the interface controls whether heat enters the layer, while thickness only sets the lateral distance once access is established. After admission, the in-layer mean free path governs the diversion---AlN (92.3~nm) forms a lateral shunt that lowers the spreading resistance by 38\% and 17\% relative to the AlGaN-capped and bare-GaN cases, whereas the short-MFP AlGaN cap (7.5~nm) traps heat and raises it by 33\%; a near-junction placement thus outperforms the far boundary.

(2) LO pumping changes both the magnitude and the ordering, raising the spreading resistance by 58--83\% in molecular dynamics and 9--20\% in full-band Monte Carlo and weakening or reversing the equilibrium optimum. A spectral conversion gate then precedes every spatial gate, locking energy in over-populated optical modes until optical-to-acoustic conversion; lacking any optical--acoustic distinction, the gray model misses this step and its three structures converge.

(3) The three gates in series---optical-to-acoustic conversion, interface admission, and lateral diversion---unify both methods and all three structures in one closed non-equilibrium resistance network. The conversion gate, governed by $\Pi=L_{\mathrm{op}\to\mathrm{ac}}/\Lambda_{\mathrm{ac}}$, contributes a saturating resistance $\eta_{\mathrm{op}}C_c\Pi/(1+\Pi)$, while a cap acts as a shunt or a trap according to its spectral transmission and diversion efficiency; the network predicts five of six Monte Carlo cases to within 1.6\% and the last to within 6.5\%.

(4) The picture gives direct design rules: accelerate conversion to keep $\Pi<1$, engineer interfaces for large laterally directed transmission, and make the cap highly conductive, with thickness and position secondary. The branch-resolved rates remain phenomenological and the methods agree on the $\sqrt{N}$ scaling rather than absolute magnitudes, so branch-resolved first-principles rates and device validation are the next steps; near-junction design thus becomes a joint spatial-and-spectral rather than scalar thermal-conductivity problem.

\begin{center}\textbf{SUPPLEMENTARY MATERIALS}\end{center}
\noindent Supplementary materials are provided as a separate document and include the NEMD protocol and interatomic potential (Note~S1), the interface models (Note~S2), the geometry and boundary conditions (Note~S3), the first-principles spectra and full-band Monte Carlo setup (Note~S4), the high-thermal-conductivity bottom-layer analysis (Note~S5), the spectral and structural analysis tools (Note~S6), the heat-source implementation (Note~S7), the parameter sweeps and factorial design (Note~S8), and the derivation and calibration of the equivalent thermal-resistance model (Note~S9), together with Figures~S1--S13 and Tables~S1--S11.

% ===================== Declarations =====================
\begin{center}\textbf{DATA AVAILABILITY STATEMENT}\end{center}
\noindent The data supporting the findings of this study are available from the corresponding author upon reasonable request.

\begin{center}\textbf{COMPETING INTEREST}\end{center}
\noindent The authors declare no competing interests.

\begin{center}\textbf{ACKNOWLEDGMENTS}\end{center}
\noindent This work was financially supported by the National Natural Science Foundation of China (Nos.~52206105 and 52425601).

% ===================== References =====================

\end{document}